**Title:**

# Micro-pixelated halide perovskite photodiodes fabricated with ultraviolet laser scribing

**Authors:**

A.P. Morozov[1], P.A. Gostishchev[1], A. Zharkova[1], A.A. Vasilev[2], A.E. Aleksandrov[3], A.R. Tameev[3], A.R. Ishteev[1], S.I. Didenko[2] and D.S. Saranin[1*]

**Affiliations:**

[1]LASE – Laboratory of Advanced Solar Energy, NUST MISiS, 119049 Moscow, Russia

[2]Department of semiconductor electronics and device physics, NUST MISiS, 119049 Moscow, Russia

[3]Laboratory "Electronic and photon processes in polymer nanomaterials", Russian Academy of Sciences A.N. Frumkin Institute of Physical chemistry and Electrochemistry, 119071, Moscow, Russia

**Corresponding author:** Dr. Danila S. Saranin saranin.ds@misis.ru

**Abstract:**

In this study, we present a complex investigation for miniaturizing of perovskite photodiodes (PPDs) in various geometries with use of ultraviolet laser scribing (UV-LS). Employing a 355 nm (3.5 eV) pulsed laser at 30 kHz, we successfully manufactured PPDs with pixel configurations of 70×130 μm², 520×580 μm², and 2000×2000 μm². The utilization of UV-LS has a proven efficiency in achieving relevant diode characteristics, such as low dark currents and high shunt resistance, as well as ultrafast response. The multi-step scribing cycle provided precise patterning of perovskite photodiodes (PPDs) in a string design. The dark current densities demonstrated exceptional uniformity, ranging from $10^{-10}$ A/cm² for 2000x2000 μm² pixelated PPDs to $10^{-9}$ A/cm2 for the 70x130 μm² configuration. The string PPDs, consisting of 10 pixels per string, displayed homogenous dark current values, ensuring effective isolation between devices. Under green light illumination (540 nm), all PPD types exhibited a broad Linear Dynamic Range (LDR). Specifically, LDR values reached 110 dB, 117 dB, and 136 dB for 70x130, 520x580, and 2000x2000 devices, respectively, spanning an illumination intensity range from $2·10^{-3}$ mW/cm2 to 2 mW/cm2. High responsivity values up to 0.38 A/W, depending on the PPDs' geometry, highlight the potential of laser scribing devices for sensing in the visible range. The calculated specific detectivity performance (from $10^{11}$ to $10^{13}$ Jones) surpasses commercial analogs, while the sub-microsecond response of 70x130 μm² and 520x580 μm² miniaturized devices underscores their suitability for precise time resolution detection systems.

**Introduction:**

Photodiode arrays are essential components of various sensor systems, including imaging[1], communication[2], spectroscopy[3], etc.[4]. The miniaturization of photodiodes (**PDs**) is critical for optoelectronic technologies and offering multiple benefits[5]. Downsizing sensing components enables the development of portable devices and increase pixel density[6]. The reduced electrical capacitance in miniaturized photodiodes leads to faster response times and lower power consumption[7]. Smaller physical dimensions of the device mitigate the impact of the noise effects[8]. This allows for more precise detection of weak optical signals, particularly in low-light conditions.

In recent years, there has been significant research focusing on the application of halide perovskites (HPs) in optoelectronics, particularly in PDs[9,10]. The chemical composition of halide perovskites is represented by the formula $ABX_3$, where A-cation is organic cation formamidine ($FA^+$) or cesium ($Cs^+$); B-cation is typically lead ($Pb^{2+}$), and X-anion is iodine (I), bromine ($Br^-$) or chlorine ($Cl^-$)[11]. Halide perovskites (**HPs**) have strong optical absorption in the visible region ($>10^5$ cm$^{-1}$)[12], tunable band-gap (from 1.2 to 2.8 eV)[13,14], high values of lifetimes (up to microseconds)[15] and reduced dynamics of non-radiative recombination[16]. Typically, HP-based photodiodes are thin-film structures with nanocrystalline absorber. The possibility of fabrication with solution processing (slot-die coating[17], inkjet printing, etc.) potentially allows cost-effective mass production. Various photodiode operating scenarios require different degrees of miniaturization, with device areas ranging from units to less than $10^{-3}$ square millimeters (mm$^2$). For instance, for X-ray flat panel detectors based on amorphous silicon technology, the pixel pitch can range from around 100 μm to 200 μm or more[18]. The technological cycles of photodiode production for classical semiconductors (Si, $A_3B_5$, etc.) are based on the use of photolithography processes[19,20], which ensure accuracy and reproducibility. However, the application of photolithography to HP thin-films faces challenges due to the nature of perovskite materials. Halide perovskites can be sensitive to solvents and high-energy UV light, which can lead to the decomposition of perovskite absorber layer and the complicated lift-off processes[21]. Patterning perovskite films and functional layers using photolithography requires careful consideration of the perovskite's sensitivity to solvents and the need for sacrificial layers[22,23]. Hence, a significant challenge in perovskite photodiodes involves developing alternative micron-scale patterning techniques. Ultraviolet (UV) laser scribing for patterning halide perovskite devices offers several advantages compared to other methods. The technology operates by exploding material rather than melting it, which produces a cleaner edge and less thermal impact. The laser scribing method is a fast, facile, and non-contact process that provides precise and intricate patterning of perovskite materials. Wide-bandgap materials, perovskite absorbing layers, and metal films can be processed using ultraviolet scribing thanks to its high photon energy. Currently, extensive experience in UV-LS has been gained for patterning solar modules using multi-stage P1-P3 processing[24–26]. However, the application of laser ablation for miniaturizing perovskite PDs requires further research and validation.

In this study, we present a comprehensive investigation of the application of UV laser scribing for miniaturizing perovskite photodiodes in various geometries. Using a 355 nm (3.5 eV) pulsed laser with an operating frequency of 30 kHz, we fabricated photodiodes with active areas ranging from ~4.00 to ~0.009

mm$^2$. The utilization of UV-LS has a proven efficiency in achieving relevant diode characteristics (low dark currents and high shunt resistance), as well as fast response. We confirmed the possibility for the fabrication of micro-pixelated photodiodes in string form. The obtained results we deeply analyzed and discussed.

**Results and discussion**

We used on a p-i-n architecture of the PDs with the following stack: glass (1.1 mm)/ITO (anode, 330 nm)/NiO (p-type, 30 nm)/perovskite absorber $Cs_{0.2}FA_{0.8}PbI_{2.93}Cl_{0.07}$ (450 nm)/PCBM (n-type, 30 nm)/BCP (hole blocking interlayer, 10 nm)/Copper (cathode, 100 nm). $Cs_{0.2}FA_{0.8}PbI_{2.93}Cl_{0.07}$ has a bandgap width of 1.58 eV with an absorption edge at 785 nm[27]. The use of the perovskite absorber $Cs_{0.2}FA_{0.8}PbI_{2.93}Cl_{0.07}$ enables strong photoelectric conversion in the visible range, with an external quantum efficiency for the device of over 80%[27]. A detailed description of the experimental section presented in supplementary information (**S.I.**). The formation of the p-i-n photodiode structure involved the multistage deposition of the functional layers and laser ablation with a specific pattern. The general sequence of technological processes is as follows:

1. Scribing of the ITO electrode (P1) with formation of the isolation areas.
2. Deposition p-i-n device stack (NiO, perovskite absorber, PCBM, BCP).
3. Scribing of p-i-n stack in alignment to the ITO electrode geometry (P2).
4. Deposition of the back metal electrode.
5. Ablation of the metal from the electrical insulation areas/ Scribing of the metal to form the photodiode pixel (P3).

To perform various miniaturization of PPDs, we aimed to achieve pixel geometries of 70×130 μm$^2$, 520×580 μm$^2$, and 2000×2000 μm$^2$ with pixel area of $9.1·10^{-5}$ cm$^2$, $3.0·10^{-3}$ cm$^2$ and $4.0·10^{-2}$ cm$^2$, respectively. General schematics of the photodiode structure patterning sequence presented on the **fig.1(a)(b).** The active area of the photodiode pixel is the overlapping area of the anode (ITO) and back electrode (copper) planes (**fig.1(c)**). ITO scribing was performed using 3 W power at a rate of 5 mm/s (50 kHz, 1 pulse per 3 microns). Electrical isolation between the anode electrodes of the ITO for each pixel in the row was achieved by sequentially conducting 9 passes of the laser beam (50-micron diameter) with an offset of 10 microns. The width of the electrical isolation area across the ITO coating was 130 μm. Electron microscopy images after P1 scribing are attached in **Fig. S1-S3** in **S.I.**. The average widths of the anode electrodes (± standard deviation) were 70.5 μm (±1.9 μm), 521 μm (±2.2 μm), and 1998 μm (±3.5 μm). After P1 scribing on ITO substrates, the profile of the scribing lines was investigated to identify possible cladding during localized overheating of the material. Stylus profilometry data showed minor surface variations, with ±4.2 nm, when comparing the areas of the cut edge and the center of the substrate. After layer-by-layer deposition of the p-i-n device architecture, P2 scribing was performed with full alignment to the P1 pattern.

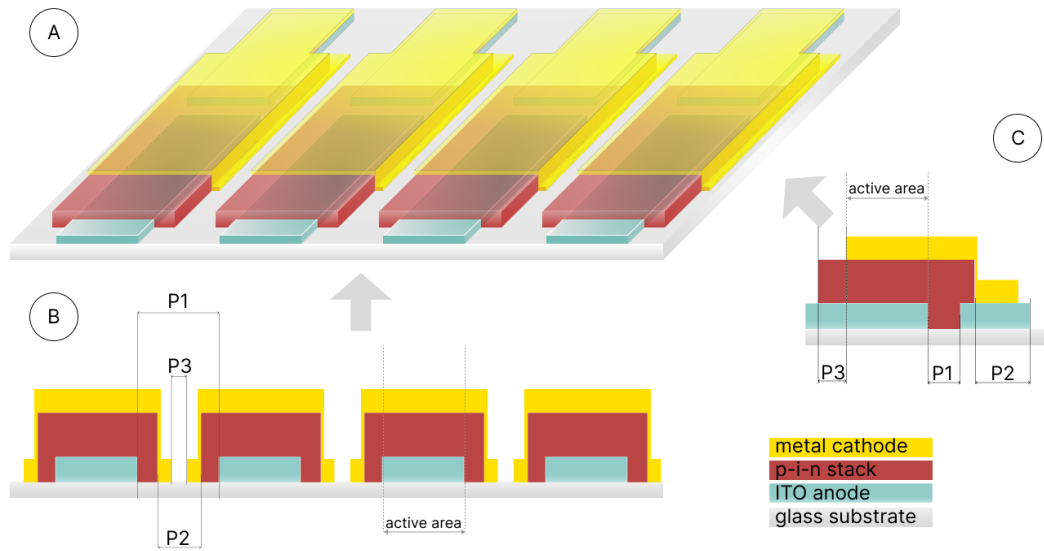

Figure 1 — Schematic image of laser scribing path P1 – ITO anode scribing, P2 — p-i-n stack scribing, P3 – metal cathode scribing: a) isometric view, b) front view and c) profile view

The P2 process was realized in three passes (5 μm offset) at 1 W power at a speed of 5 mm/s. The width of the scribing line of the P2 process was ~60 μm. After deposition of the metal electrode in a vacuum, the P3 process required the removal of conductive material from the insulating zones between the ITO anode electrodes. Additionally, transverse scribing of the metal electrode was performed to form the final pixel geometry. The SEM image in **Fig. 2(a)** shows a string of 10 PPD pixels. The pixel formation, indicated by the transverse metal scribing line in Fig. 2(a), required the process to be performed at 1 W power, a speed of 2 mm/s, and three passes (10 μm offset). The width of the cut was 40 μm. Electrical isolation between pixels was achieved by ablating the metal contact using the P1 path (side isolation lines according to the top view in Fig. 2(a)). To ensure isolation, scribing was performed at a power of 3 W, speed of 20 mm/s for 1 pass of the laser beam with a cutting width of 40 μm. The P1-P3 patterning processes were the same for PPDs of all configurations. Analysis of SEM images of PPDs (**fig.2 (b)(c)(d)**) after the P1-P3 processes enabled us to estimate the pixel lengths. The length values (± standard deviation) of 132.3 μm (±2.9 μm), 576.2 μm (±3.6 μm), and 1998.3 μm (±13.6 μm) were obtained for the PPD configurations fabricated with different levels of miniaturization and were consistent with the target values. A close inspection of the SEM image in Fig. 2(b) shows edge non-uniformity during the scribing of the Cu metal electrode. After the P3 process, we observed partial edge delamination of the cathode and areas of jagged shape along the laser ablation line. We attribute this effect to possible mechanical stresses between the metal film and the pin stack, as well as to the insufficient attenuation of the laser beam needed to reduce local thermal effects. However, the area of inhomogeneity did not exceed 15 μm relative to the pixel edges. Therefore, assessing the impact of this technological imperfection requires a comprehensive analysis of the device performance.

To simplify the naming of PPD configurations in this work, we will use the abbreviations 70x130, 520x580, and 2000x2000 to denote their sizes. Photo–images of the resulting strings of PPDs for different miniaturizations are presented in **Figs. S4-S6** in the **S.I.**

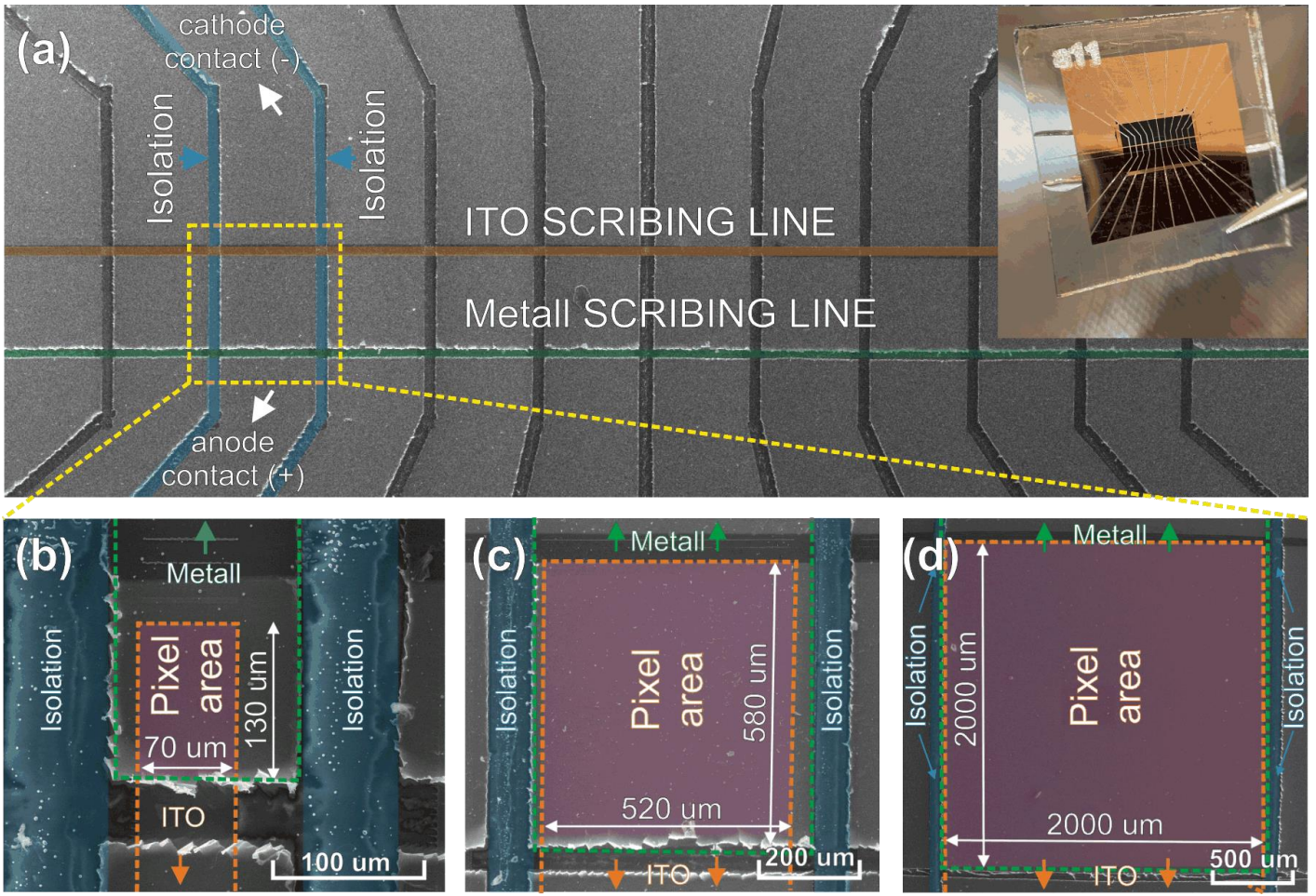

Figure 2 — SEM images of the string with PPDs (520x580) with inset of photo-image (a) SEM image of the pixel 70x130 (b); SEM image of the pixel 520x580 (c) and 2000x2000 (d)

To investigate the effect of device geometry on diode properties, we measured the dark volt-ampere characteristics for a single PPD, as shown in **Fig.3**. A comparative analysis was performed for the current density-voltage curves (JV curves). The measurements were taken from 1.1 V for the positive voltage to 0.3 V for the negative one. The dark JV curves for all types of PPDs exhibited typical diode behavior, characterized by four main regions related to shunt current (I), recombination current (II), diffusion current (III), and contact resistance (IV). The minimum value of the dark current density ($J_D$) at a voltage of ~0 V exhibited an inverse proportionality to the device area. Specifically, for the 70x130 PPD, $J_D$ was $4.9 \times 10^{-9}$ A/cm$^2$; for 520x580 PPD $J_D = 6.5 \times 10^{-10}$ A/cm$^2$, and for the 2000x2000 device $J_D = 1.4 \times 10^{-10}$ A/cm$^2$. The increase in dark current for smaller PPDs may result from the influence of edge effects. Laser ablation process can induce local decomposition of the perovskite and the formation of micro-shorts along the boundaries of the device structure. The overheating of hybrid perovskites containing leads to the formation of metallic lead[28,29], which could contribute to the current leakage or recombination dynamics. As reported in works [24,30,31], generation of the metal dust during the scribing of metal contact provides shunting pathways at the edges of the device structure. The calculation of shunt resistance ($R_{sh}$) revealed a decrease in value from $1.48·10^7$ Ohm·cm$^2$ for the 2000x2000 PPD to $4.76·10^5$ Ohm·cm$^2$ for 70x130 device. The dark JV curves changed the form with the alteration in PPD geometry. Specifically, the device with maximum miniaturization (70x130) exhibited an increased recombination current contribution (**fig. 3(a)**), while 520x580 and 2000x2000 configurations showed a sharp increase in diffusion current (**fig. 3(b)(c)**). The PPD with the maximum area (2000x2000 um$^2$) had the shift of minimal dark current point from 0 to 0.07 V. This effect can be attributed to dark hysteresis in p-i-n architectures based on halide perovskites with NiO hole-transporting layer. Specifically, the appearance of the built-in potential in dark JVs related to the migration of ionic defects formed at the chemically active NiO/perovskite interface[32,33]. Also, we measured the dynamics of $J_D$ stabilization with time (**fig.S7 in S.I.**). PPD 70x120 and 520x580 showed stable output without saturation effects, while the device 2000x2000 demonstrated a dynamic decrease from during 30 s before reaching constant values.

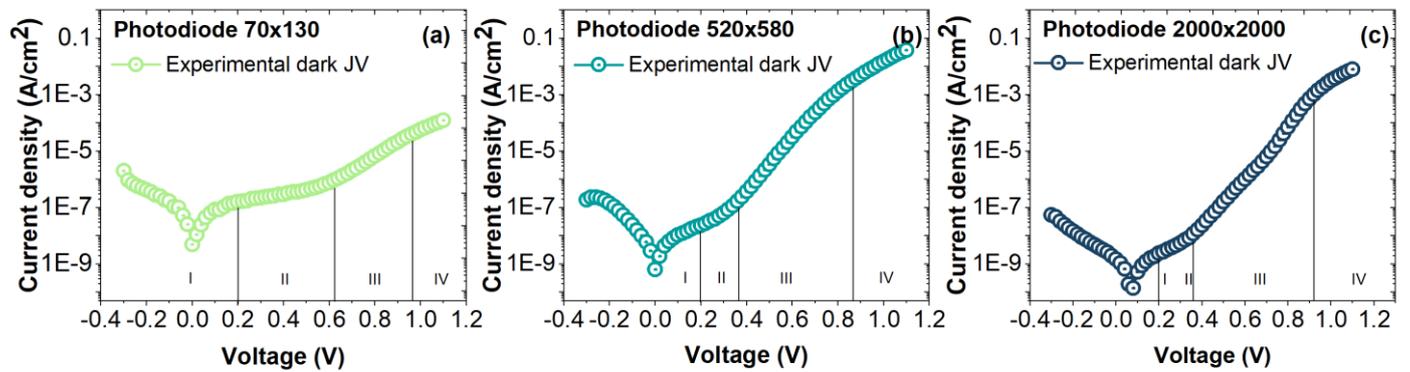

Figure 3– Dark JV curves with fitting by 2-diode model for 70x130 um$^2$ photodiode (a); 520x580 um$^2$ photodiode and 2000x2000 um$^2$ photodiode

The analysis of $J_d$ homogeneity in a string of 10 pixels revealed differences in values within the one order of magnitude for all configurations (histograms presented in **fig. S8 in S.I.**). For the 70x130 PPDs string, the mean value (standard deviation) was - $6.2·10^{-9}$ A/cm$^2$ (±$1.3·10^{-9}$ A/cm$^2$), for the 520x580 configuration - $7.4·10^{-10}$ A/cm$^2$ (±$8.1·10^{-11}$ A/cm$^2$), and for the 2000x2000 devices - $2.2·10^{-10}$ A/cm$^2$ (±$4.4·10^{-11}$ A/cm$^2$). The photo-response of the PPD to different light intensities was estimated for a wide range of illumination power density ($P_0$, light source - green LED, 540 nm) from ~$2·10^{-3}$ mW/cm$^2$ to ~2 mW/cm$^2$. The dependencies of the open circuit voltage ($V_{oc}$) and short circuit current density to $P_0$ presented on the **fig.4**. All PPD configurations showed a logarithmic trend of $V_{oc}$ increase illumination power density (**fig.4(a)**), while for $J_{sc}$ we obtained linear dependencies (**fig.4(b)**). As the $P_0$ increases linearly, the $J_{sc}$ follows the trend, since photocurrent $\propto$ carrier generation rate[34]. The $V_{oc}$ increases logarithmically due to the limitation of the contact potential $V_{oc} \propto$ ln (photocurrent/dark saturation current) [34]. At minimum $P_0$ values of ~$2·10^{-3}$ mW/cm$^2$, all PPD configurations exhibited close $V_{oc}$ values ranging from 0.32 to 0.35 V. As $P_0$ increased, the $V_{oc}$ growth dynamics of the 70x130 PPD differed significantly from those of the 520x580 and 2000x2000 geometries. The maximum $V_{oc}$ values for the 70x130 device at 2.04 mW/cm$^2$ were 0.657 V, while the 520 and 580 PPD achieved 0.896 V and the 2000x2000 PPD achieved 0.899 V. At $P_0$~$10^0$ mW/cm$^2$, the $V_{oc}$ values for the 2000x2000 PPDs reached saturation. For miniaturized PPDs geometries, we observed less rapid increase at low $P_0$ and stronger increment and high $P_0$. The photocurrent response to different illumination power densities resulted in an almost linear increase from $3.2 \cdot 10^{-7}$ to $1.7 \cdot 10^{-4}$ A/cm$^2$ for the 70x130 PPD, from $4.0 \cdot 10^{-7}$ to $4.7 \cdot 10^{-4}$ A/cm$^2$ for the 520x580 PPD and from $7.2 \cdot 10^{-7}$ to $8.2 \cdot 10^{-4}$ A/cm$^2$ for the 2000x2000 PPD. The slope coefficients at linear fitting (k) were 1.019 for the 520x580 device and 1.027 for the 2000x2000. The 70x130 photodiodes exhibited reduced k= 0.932. The evaluation of the linear dynamic range (LDR) (**eq. S1** in **S.I.**) revealed values of 136 dB for the 2000x2000 PPD, 117 dB for the 520x580 PPD, and 110 dB for the 70x130 device. The decrease in LDR for 520x580 devices relative to the 2000x2000 configuration is caused by the increase in dark current. The 70x130 PPD showed an increase in $J_D$ by almost an order of magnitude compared to devices with a larger pixel size. We calculated the values of the responsivity (**R**, **eq. S2 in S.I.**) for a wavelength of 540 nm. The values were 0.011 A/W for a 70x130 device, 0.22 A/W for 520x580, and 0.38 A/W for 2000x2000. Comparison of the data obtained with the current literature[35] suggests relevantly high R values for conversion of green light.

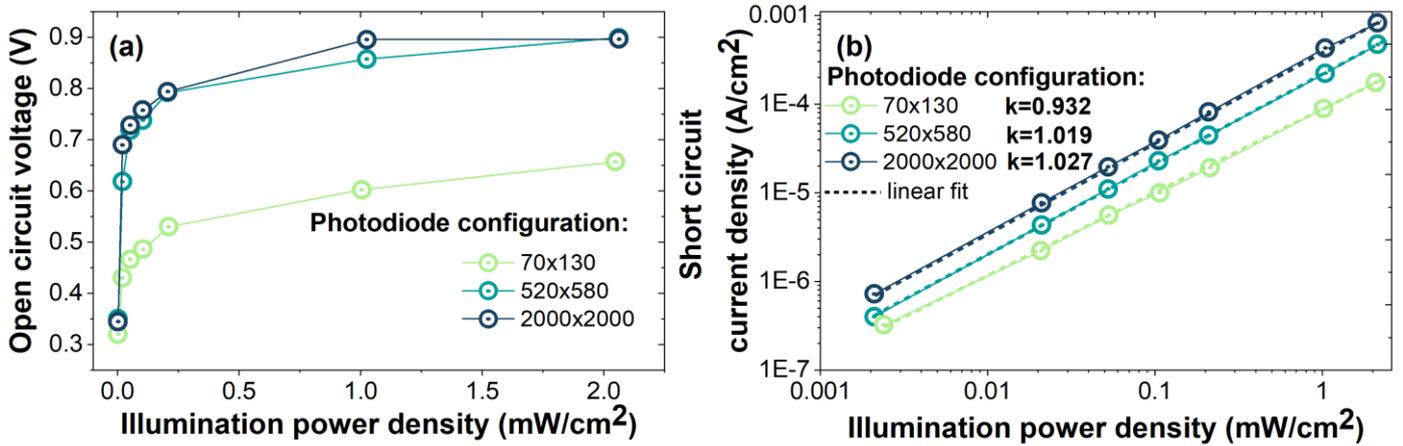

Figure 4 – The dependence of $V_{oc}$ to $P_0$ for the various geometries of PPDs (a); the linearity plot of $J_{sc}$ vs. $P_0$ for various geometries of PPDs (b)

Following the assumption of predominating shot-noise in PPDs[36], we estimated key parameters of the devices- specific detectivity (**D\*** at 540 nm) and noise equivalent power (**NEP**) at 0V. Accord to the equations **S3, S4**. calculated values of D* and NEP for the fabricated PPDs presented on the **tab.1**.

Table 1 – Calculated D8 and NEP for the fabricated PPDs

| PPD geometry | D* (Jones) at 540 nm | NEP (W·Hz$^{-1/2}$) |
|---|---|---|
| 70x130 | $6.72 \cdot 10^{11}$ | $1.42 \cdot 10^{-14}$ |
| 520x580 | $4.82 \cdot 10^{12}$ | $1.14 \cdot 10^{-14}$ |
| 2000x2000 | $1.81 \cdot 10^{13}$ | $1.10 \cdot 10^{-14}$ |

The reduced efficiency in collecting photocarriers and the weakened shunt properties due to the miniaturization of the photodiode (previously evaluated by LDR and $I_{ph}$ vs. $P_0$) clearly affected the specific detectivity. As the device sizes were reduced, D* decreased by two orders of magnitude from $1.8 \cdot 10^{13}$ Jones (2000x2000) to $6.7 \cdot 10^{11}$ Jones (70x130). The NEP values for all PPD configurations were in the same order of magnitude, around $10^{-14}$ W·Hz$^{-1/2}$. For the 2000x2000 PPD, the value decreased by approximately 30% relative to the 70x130 geometry. Notably, The D* values for the fabricated devices in this study are comparable to commercial silicon, GaP photodiodes for the visible range (D*~$10^{12}$ Jones[37,38]) and the most efficient PPDs reported in the literature[39–41].

To evaluate the dynamic performance of the photodiodes, we measured the transient photocurrent during optical pumping (**Fig. 5**). We used a green LED with a square light pulse to estimate the rise and fall time of the signal ($t_r$ and $t_f$, respectively). The transition time values were derived from the difference between the corresponding 10% and 90% saturation current values. The 70x130 and 520x580 PPDs exhibited a distinct signal waveform at 50 kHz, whereas the 2000x2000 device displayed a relevant profile at 10 kHz. Devices with the smallest area exhibited maximum performance due to the photodiode's reduced geometric capacitance. While the 70x130 device achieved a sub-microsecond response time with $t_r$=47 ns and $t_f$=45 ns, the 520x580 PPD experienced longer transient times of 98 and 561 ns. The 2000x2000 PPD had $t_r$ of 2.88 μs and $t_f$ of 3.693 μs. The dynamic characteristics of PPDs of all configurations showed an asymmetry in signal rise and fall, leading to an increase in $t_f$ relative to $t_r$. We attribute this effect to the contribution of ionic defect migration, which is characteristic of thin-film p-i-n oriented architectures based on halide perovskites[42–44].

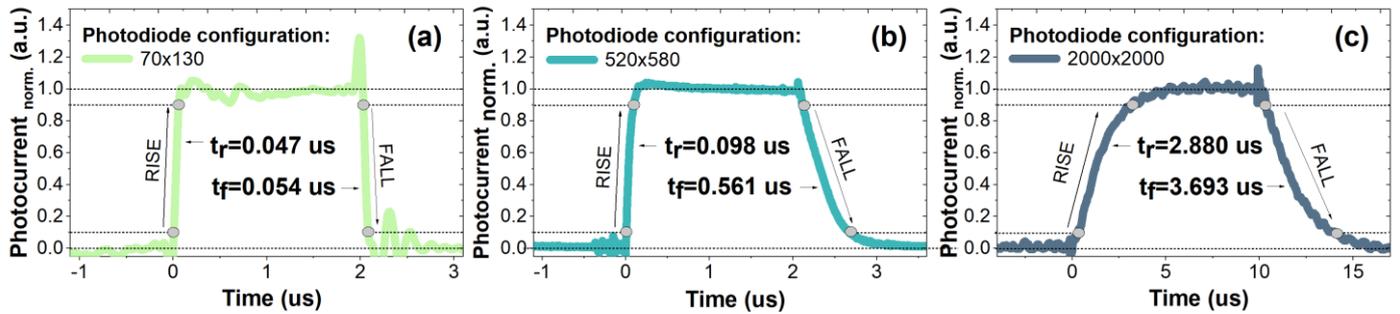

Figure 5 – The dynamic response of PPD with various miniaturization: 70x130 device (a); 520x580 device (b); and 2000x2000 (c)

Nevertheless, the miniaturized PPDs exhibited ultrafast sub-microsecond response superior to analogues based on standard amorphous silicon technologies[45,46]. To the best of our knowledge, our approach to miniaturizing photodiodes has allowed us to achieve a response speed comparable to the most efficient PPDs fabricated using lithographic techniques[39,47].

In conclusion, the application of UV-LS for PPD miniaturization is an effective approach for achieving state-of-the-art performance for visible light detection. The multi-step P1-P3 scribing cycle satisfies the requirements for obtaining PPDs of a specific shape tailored to the various detector systems currently in use. The technological imperfections, such as partial delamination of the metal electrode and non-uniform cutting edge, which influenced the output parameters of the designed PPDs, were identified. We observed that the negative impact of edge effects was significant for devices with the smallest pixel geometry (70x130), while for 520x580 and 2000x2000 configurations, the effect was not critical. The dark current densities for the 2000x2000 pixelated PPDs were on the order of $10^{-10}$ A/cm$^2$ and didn't exceed $10^{-9}$ A/cm$^2$ when the size was reduced to 70x130. High homogeneity of dark current values was achieved for the fabricated string PPDs with 10 pixels per string, ensuring complete isolation of the devices from each other. PPDs of all types displayed a broad LDR when optically pumped with green light (540 nm). The LDR was 110 dB for 70x130 devices, 117 dB for 520x580, and 136 dB for 2000x2000 over the illumination intensity range from $2 \cdot 10^{-3}$ mW/cm$^2$ to 2 mW/cm$^2$. High responsivity values, ranging from 0.011 to 0.38 A/W, depending on the PPDs' geometry, highlight the potential of laser scribing devices for sensing in the visible range. The miniaturized PPDs developed in this work can potentially provide an effective solution for converting of scintillating light from cesium iodide (CsI) or Cerium-doped gadolinium aluminum gallium garnet (GGG) sources in medical diagnostics and safety applications. The achieved D* performance of $10^{11}$ to $10^{13}$ Jones is state-of-the-art compared to commercial analogs. Moreover, the sub-microsecond response of the 70x130 and 520x580 miniaturized devices demonstrates the high potential of PPDs for precise time resolution detection systems. We believe that the further progress in the development of simplified PPD technology, in the form of strings and matrix arrays, requires new technological approaches for edge insulation of the single pixels, as well as the optimization of laser processing.

## Acknowledgments

The authors gratefully acknowledge the financial support from Russian Science Foundation with project № 21-19-00853.

## Author contributions

**D.S.S. and S.I.D.** conceived the work.

**A.P.M, P.A.G, A.Z. and A.R.I.** performed the experiments on PPDs and the electro characterization.

**A.E.A. and A.R.T.** performed measurements of the dynamic response.

**D.S.S. and S.I.D.** provided administrative support and resources.

**S.I.D.** coordinated the research activity.

The manuscript was written with contributions from all the authors. All the authors approved the final version of the manuscript.

# The supplementary information for the paper:
# Micro-pixelated halide perovskite photodiodes fabricated with ultraviolet laser scribing

**Authors:**

A.P. Morozov[1], P.A. Gostishchev[1], A. Zharkova[1], A.A. Vasilev[2], A.E. Aleksandrov[3], A.R. Tameev[3], A.R. Ishteev[1], S.I. Didenko[2] and D.S. Saranin[1*]

**Affiliations:**

[1]LASE – Laboratory of Advanced Solar Energy, NUST MISiS, 119049 Moscow, Russia

[2]Department of semiconductor electronics and device physics, NUST MISiS, 119049 Moscow, Russia

[3]Laboratory "Electronic and photon processes in polymer nanomaterials", Russian Academy of Sciences A.N. Frumkin Institute of Physical chemistry and Electrochemistry, 119071, Moscow, Russia

**Experimental section:**

*Materials*

All organic solvents—dimethylformamide (DMF), N-Methylpyrrolidone (NMP), isopropyl alcohol (IPA), chlorobenzene (CB) were purchased in anhydrous, ultra-pure grade from Sigma Aldrich, and used as received. Ethylacetate (EAC, 99+% purity) was purchased from Reaktivtorg-Himprocess hps, 2-Methoxyethanol was purchased from Acros Organics (99.5+%, for analysis), $HNO_3$ (70%). Photodiodes were fabricated on $In_2O_3$: $SnO_2$ (ITO) coated glass ($R_{sheet}$<7 Ohm/sq) from Zhuhai Kaivo company (China). $NiCl_2·6H_2O$ (from ReaktivTorg 99+% purity) used for HTM fabrication. Lead Iodide (99.9%), Cesium iodide (99.99%), Cesium chloride (99.99%) trace metals basis from LLC Lanhit, Russia and formamidinium iodide (FAI, 99.99% purity from GreatcellSolar), were used for perovskite ink. [6,6]-Phenyl-C61-butyric acid methyl ester (99% purity) was purchased from MST NANO (Russia). Bathocuproine (BCP, >99.8% sublimed grade) was purchased from Osilla Inc. (UK) and used for the fabrication of hole blocking layer.

*Inks preparation*

For the preparation of composition $Cs_{0.2}FA_{0.8}PbI_{2.93}Cl_{0.07}$ perovskite ink, we used CsCl, CsI, FAI, $PbI_2$ powders in a 0.07:0.13:0.8:1 molar ratio. The resulting mixture was dissolved in a DMF:NMP (volume ratio 640:360) with a concentration of 1.35 M and stirred at a temperature of 50 °C for 1 h. PCBM was dissolved in CB at a concentration of 27 mg/ml and stirred for 1 h at a temperature of 50 °C. BCP was dissolved in IPA at a concentration of 0.5 mg/ml and stirred for 8 h at a temperature of 50 °C. Before use, all solutions were filtered through 0.45 μm PTFE filters.

*Device fabrication*

Perovskite solar cells were fabricated with inverted planar architecture ITO/c-$NiO_x$/perovskite ($Cs_{0.2}FA_{0.8}PbI_{2.93}Cl_{0.07}$)/PCBM/BCP/Cu. Firstly, the patterned ITO substrates were cleaned with detergent, de-ionized water, acetone, and IPA in the ultrasonic bath. Then, substrates were activated under UV-ozone irradiation for 30 min. $NiCl_2·6H_2O$+$HNO_3$ precursor for NiO HTM film was spin-coated at 4000 RPMs (30 s), dried at 120 °C (10 min), and annealed at 300 °C (1 h) in the ambient atmosphere. Perovskite absorber film was crystallized on the top of HTM with solvent engineering method. Perovskite precursor was spin-coated at 3000 RPMs (5 s), and 5000 RPMs (30 s), 200 μL of EAC were poured on the substrate on the 21st second after the start of the rotation process. Then, substrates were annealed at 85 °C (1 min) and 105 °C (30 min) for conversation into the black perovskite phase. The PCBM ETL was spin-coated at RPMs (30 s) and annealed at 50 °C (5 min). BCP interlayer was also spin-coated at 4000 RPMs (30 s) and annealed at 50 °C (5 min). The copper cathode was deposited with the thermal evaporation method at 2 × 10⁻⁶ Torr vacuum level. All devices were encapsulated with UV epoxy from Osilla inc. UV LS processes (P1-P3) were described in the manuscript.

*Laser scriber*

The laser scriber system was designed by LLC Nordlase (Russia).

Laser type – Nd:YVO$_4$, 355 nm, impulse – 22 ns at 50kHz. Maximum power – 3W.

The positioning of the samples was realized using motorized XY stage from Standa (1 um resolution in XY movement).

The maximum attenuation of the system – 99%.

During scribing all substrates were fixed with vacuum chuck.

*Characterization*

Surface roughness and film thicknesses were measured with KLA-Telencor stylus profilometer.

SEM images were taken in high vacuum at 15 kV accelerating voltage were done with a JEOL JSM7600F system (Tokyo, Japan).

Dark JV curves were measured in an ambient atmosphere with Agilent B1500A Semiconductor Device Analyzer (voltage step of 20 mV).

We used LM1-EPG1-11-N2-00001 CREE LED (540 nm) as a light source for estimation of $V_{oc}$, $J_{sc}$ vs. $P_0$. The output parameters ($V_{oc}$, $J_{sc}$) were extracted from JV curves measured with Keithley 2400 SMU in 4-wire mode and a settling time of $10^{-2}$ s. The LED was connected to a Keysight E36311A source-meter in a hinge-mounted configuration. Optical power measurements were performed on a ThorLabs S425C. Illuminance measurements were performed on a UPRtek MK350. Completed set up for characterization of PPDs was placed in a black box.

The dynamic response was measured with DIGILENT Analog Discovery Pro 3450 (2 units), which were used as oscilloscope and pulse generator.

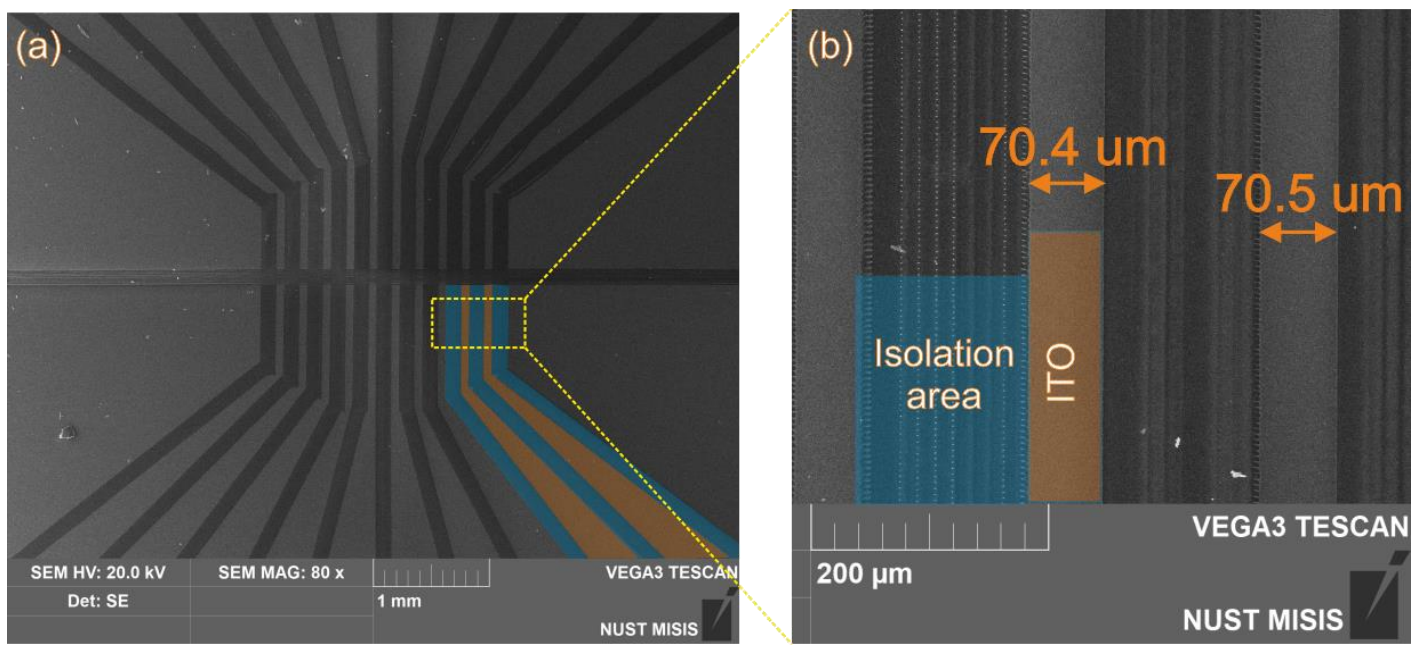

Figure S1 – The SEM image of ITO/Glass scribing after P1 process with 70 um width of anode electrode

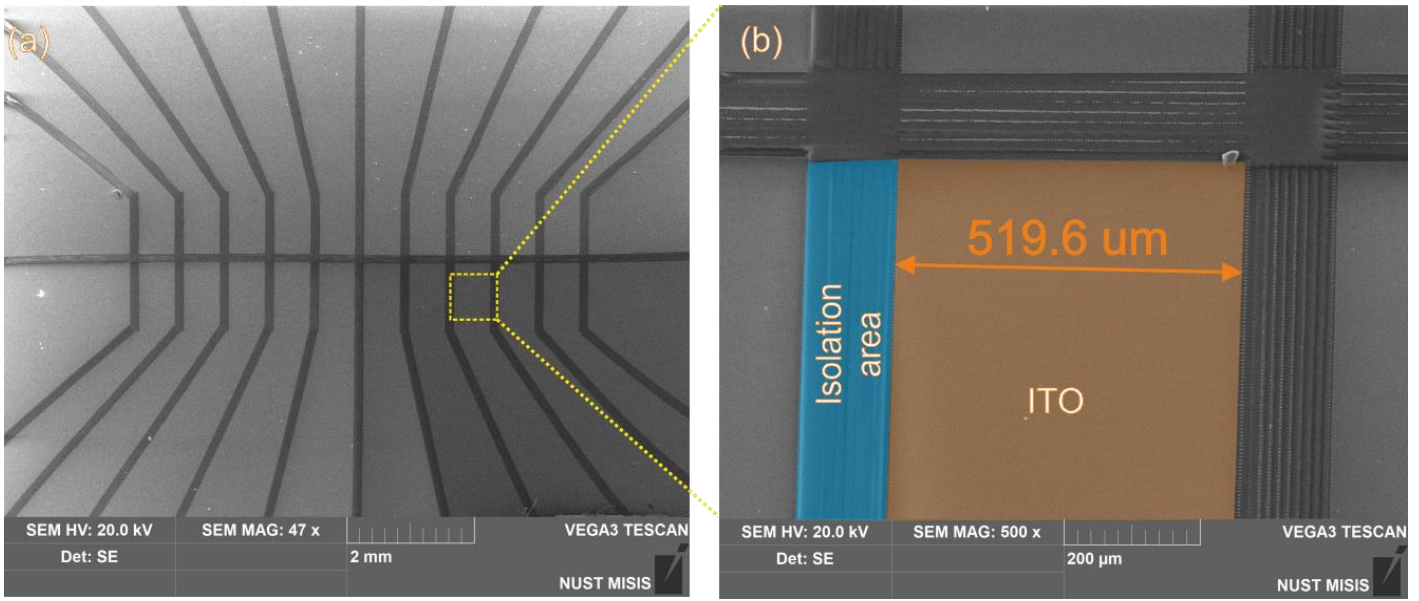

Figure S2 – The SEM image of ITO/Glass scribing after P1 process with 520 um width of anode electrode

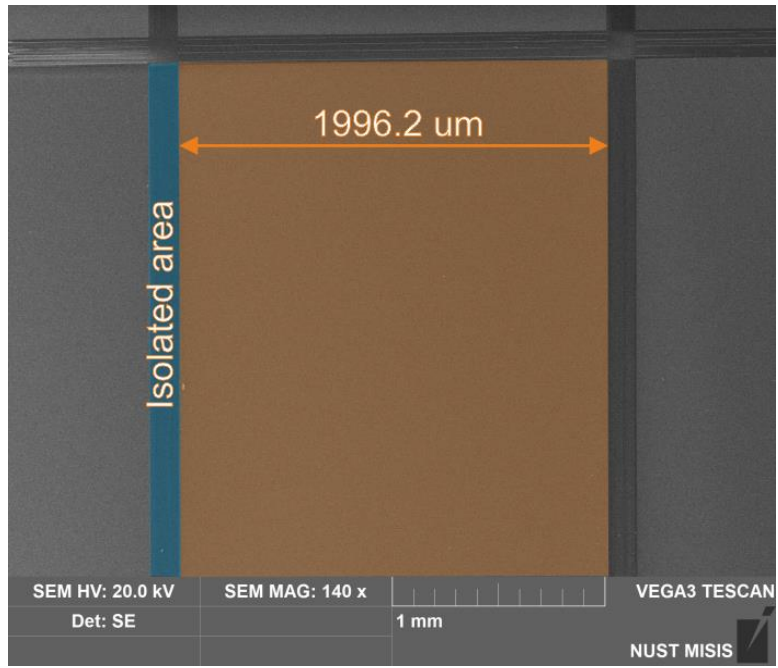

Figure S3 – The SEM image of ITO/Glass scribing after P1 process with 2000 um width of anode electrode

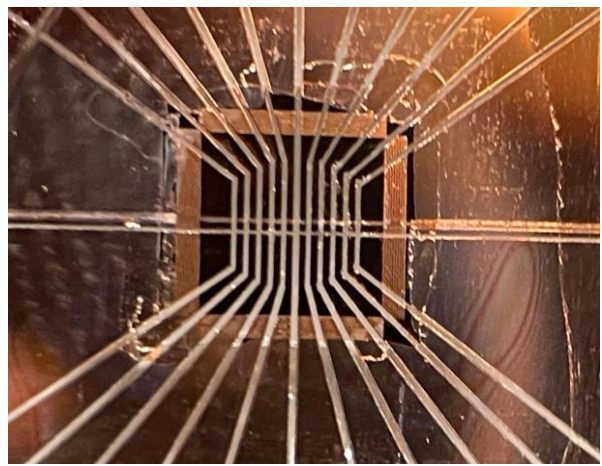

Figure S4 – Photo-image for the string of 10 PPDs with 70x130 um$^2$ pixels

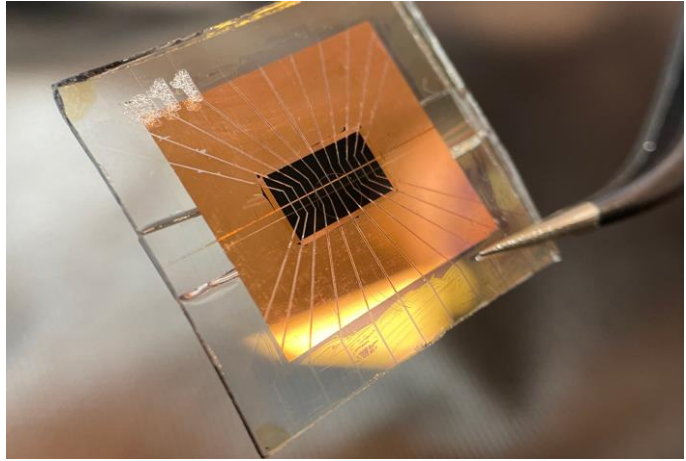

Figure S5 – Photo-image for the string of 10 PPDs with 520x580 um² pixels

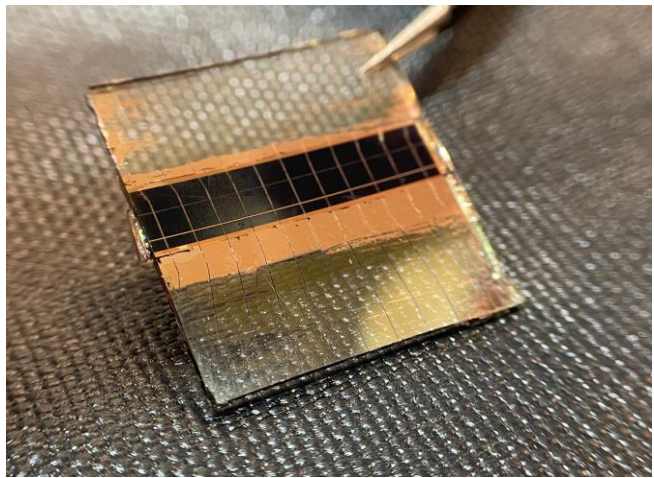

Figure S6 – Photo-image for the string of 10 PPDs with 2000x2000 um² pixels

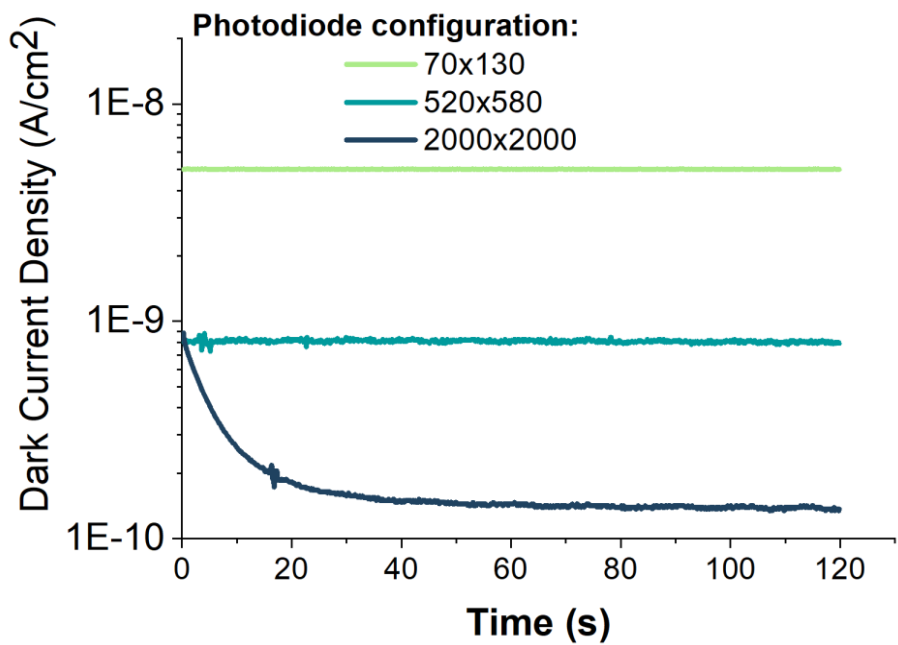

Figure S7 – The stabilization of $J_D$ for various geometries of PPDs with time

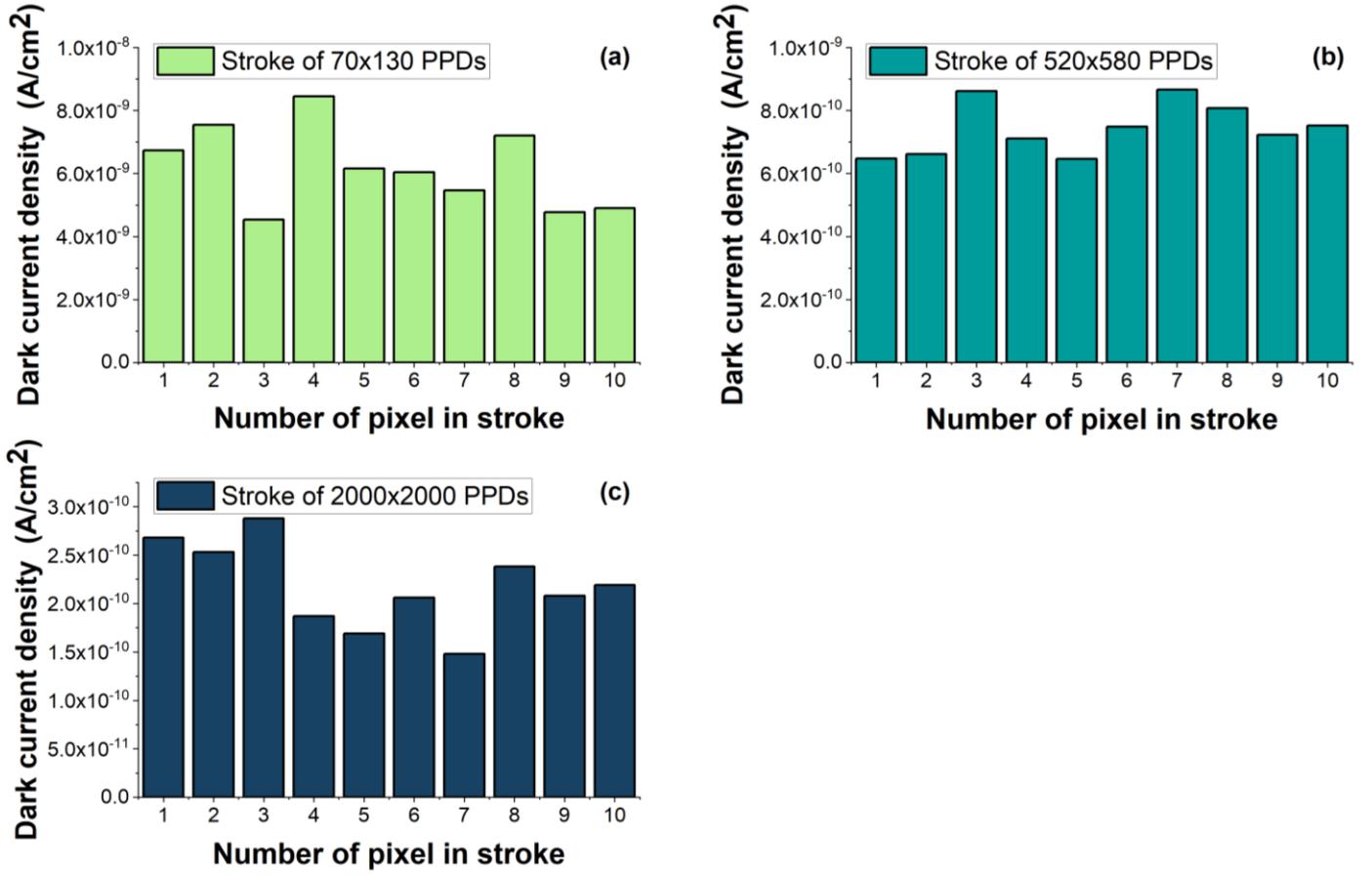

Figure S8 – The distribution of dark current density leakage values for the PPDs in stroke with geometries 70x130 (a); 520x580 (b); and 2000x2000(c)

$$LDR = 20 \log \left(\frac{I_{ph}}{I_D}\right) \quad (S1)$$

Where $I_{ph}$- photocurrent in linear range

$I_D$- dark current

$$R = \frac{I_{ph}}{P_0} \quad (S2)$$

Where $I_{ph}$- photocurrent (A)

$P_0$- power of the illumination (W)

$$D^* = \frac{I_{ph}\sqrt{A}}{P_0}\left(\frac{1}{(2qJ_d)^{1/2}}\right) \quad (S3)$$

Where A – active area of the device

$$NEP = \frac{(2qJ_d)^{1/2}}{R} \quad (S4)$$